\pdfoutput=1

\documentclass[aps,prl,amsmath,amssymb,floatfix,twocolumn,amsmath,nosuperscriptaddress,twocolumn,nofootinbib,tighten,letterpaper,10pt]{revtex4-2}

\usepackage[utf8]{inputenc}
\usepackage{bm}
\usepackage{siunitx}
\usepackage{enumerate}
\usepackage{amsfonts}
\usepackage{amsmath}
\usepackage{amssymb}
\usepackage{color}
\usepackage{soul}
\usepackage{float}
\usepackage{graphicx}
\usepackage{graphics}
\usepackage{physics}
\usepackage{verbatim}
\usepackage[colorlinks,allcolors=blue]{hyperref}
\usepackage{tikz-feynman}
\usepackage{xcolor,soul}
\usepackage{cancel}
\usepackage{tikz}
\usepackage{subcaption}
\usetikzlibrary{angles, quotes, intersections}
\usepackage[capitalise]{cleveref}
\usepackage{ragged2e} 
\usepackage{caption}
\usepackage{titlesec}

\let\epsilon=\varepsilon

\definecolor{DarkRed}{rgb}{0.80,0,0}
\definecolor{DarkGray}{rgb}{0.7,0.7,0.7}

\newcommand{\prlsection}[1]{\textit{#1}.\kern0.05em---\kern0.05em\ignorespaces}
\begin{document}

\title{Interplay of bound states in the continuum and Fano--Andreev interference in a hybrid triple quantum dot}

\author{A. Gonz\'alez I.}
\thanks{Corresponding author: alejandro.gonzalezi@usm.cl}
\affiliation{Departamento de F\'{\i}sica, Universidad T\'{e}cnica Federico Santa Mar\'{\i}a, Casilla 110 V, Valpara\'{\i}so, Chile}

\author{Pedro A. Orellana}
\affiliation{Departamento de F\'{\i}sica, Universidad T\'{e}cnica Federico Santa Mar\'{\i}a, Casilla 110 V, Valpara\'{\i}so, Chile}

\author{Vladimir Juri\v ci\'c}
\affiliation{Departamento de F\'{\i}sica, Universidad T\'{e}cnica Federico Santa Mar\'{\i}a, Casilla 110 V, Valpara\'{\i}so, Chile}


\begin{abstract}
We investigate bound states in the continuum (BICs) in a hybrid normal--superconducting triple quantum dot system, where the central dot is coupled to two normal leads and the lateral dots are proximity-coupled to superconducting electrodes. Local electron--electron interactions are treated within the Hubbard approximation. Finite bias, together with lateral-dot detuning and superconducting proximity, induces interference between elastic electron tunneling (ET) and Andreev reflection (AR) channels, mediated by BIC-related modes and proximity-induced Andreev bound states. As the bias is swept through the subgap resonances, ET exhibits sharp antiresonances that evolve into exact transport zeros, signaling the emergence of (quasi-)BICs. We further find a continuous crossover from a Fano--Andreev BIC-supported regime to a Fano--Andreev quasi-BIC regime as the detuning asymmetry increases. The formation of BICs and quasi-BICs is accompanied by a pronounced change in the occupation of the side quantum dot, providing an internal diagnostic directly correlated with the transport signatures of the bound states.
\end{abstract}

\maketitle


\emph{Introduction.} Bound states in the continuum (BICs), originally predicted by von Neumann and Wigner \cite{vN-W}, are localized states embedded in the continuum spectrum. They constitute a paradigmatic interference effect \cite{ReviewNature} and were first observed experimentally by Plotnik \textit{et al.} \cite{Plotnik}. These states have been reported in acoustic, photonic, and related fields~\cite{ReviewNature,Plotnik,Hsu2013,Huang2021}, but their experimental realization in mesoscopic electronic transport remains elusive.

A closely related interference phenomenon in mesoscopic transport is the Fano effect, which arises from the interplay between a discrete level and a continuum channel \cite{Fano.RMP,U.Fano}. It produces asymmetric resonances, antiresonances, or even complete suppression of transport \cite{PhysRevB.70.035319}. This phenomenon has been investigated in T-shaped double quantum dot systems (T-DQD) with metallic leads, both with and without a floating lead. In systems without a floating lead, the Fano effect is observed as an asymmetric line shape in the transmission spectrum, modeled as
$\mathcal{F}(\varepsilon)=\frac{(\varepsilon+q)^2}{1+\varepsilon^2}$,
where \(\varepsilon=(E-E_0)/(\Gamma/2)\), \(E_0\) is the resonance energy, \(\Gamma\) the linewidth, and \(q\) the asymmetry parameter \cite{PhysRevB.81.115316,PhysRevB.70.035319}. By contrast, an additional floating metallic lead induces decoherence and suppresses the Fano line shape \cite{Gao_Wen-Zhu_2008,PhysRevB.85.205451}. Superconducting electrodes, however, qualitatively alter this behavior.

In particular, superconducting proximity effects enable coherent electron-hole conversion via Andreev reflection, leading to Andreev bound states (ABS) inside the superconducting gap~\cite{Qing-feng-Sun, ZYu, Xiao-Qi-Wang,JanBaranski2,T.Domanski2, Yu-Zhu,ECSiqueira2,Sachin-Verma,Grzegorz-Michalek,BAI20104875,WeiPinXu,J.Gramich,TrochaPRB95,Arora2025,10-20241756,PhysRevB.111.125402}. When a superconducting lead is additionally coupled to a T-shaped double quantum dot connected to normal leads, the induced particle-hole mixing on the side dot generates ABS that interfere with the direct normal transmission channel, producing Fano-like structures in the electronic transmission between the normal leads \cite{LongBai_TQD_A,Siqueira3,CalleAM}. In this case, however, the Fano asymmetry parameter \(q\) becomes complex, so the electronic transmission no longer vanishes exactly \cite{CALLE20131474,Fano_Andreev_Effect_TQD}. Along the same line, Xu and Shi showed that a parallel double-quantum-dot structure with side-coupled superconductors can host true Andreev bound states in the continuum in the symmetric configuration, while a finite superconducting phase difference can drive their evolution into Fano antiresonances \cite{Xu2025}.

\begin{figure}[t!]
\centering
\includegraphics[scale= .8]{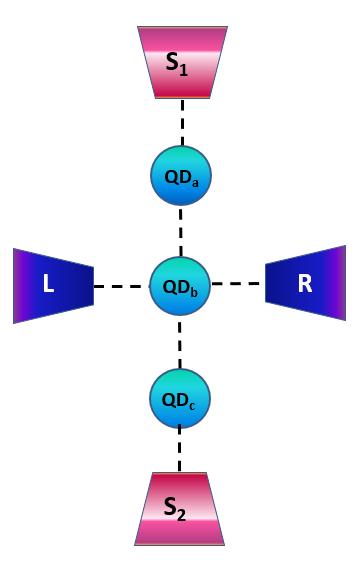}
\caption{\justifying{Schematic of the hybrid normal--superconducting triple-quantum-dot device. The central dot \(QD_b\) is coupled to the two normal leads \(L\) and \(R\), while the lateral dots \(QD_a\) and \(QD_c\) are proximity-coupled to the superconducting electrodes \(S_1\) and \(S_2\), respectively. The interdot couplings \(t_{ab}\) and \(t_{cb}\) connect the lateral dots to the central dot, and the lateral-dot levels are detuned as \(\epsilon_{a,c}=\epsilon_b\pm\eta\).}}
\label{Fig1}
\end{figure}

In comparison to the double quantum dot (DQD) structure, triple quantum dots provide additional tunability and a richer interference structure due to the extra degree of freedom. In particular, the two lateral dots support symmetric and antisymmetric combinations, which can host quasi-BICs and BICs, thereby offering a highly tunable platform for engineering and controlling such states. In the absence of additional decoherence channels, the antisymmetric combination may fully decouple from the continuum and form a BIC. However, when energy detuning is introduced between the side quantum dots, this state is no longer bound, but instead couples weakly to the continuum and becomes a quasi-BIC. These effects have been investigated in both normal and hybrid nanostructures and have been shown to strongly affect charge, spin, and thermoelectric transport~\cite{PhysRevB.87.075102, PhysRevB.57.6642,PhysRevB.70.233315,Trocha_2008,PhysRevB.87.075102,TrochaPRB78,VernekPRB82,OrellanaPRB74,OrellanaPRB73,Dicke_Fano_Andreev,Trocha2025,PhysRevB.95.125419}.

While BIC physics in triple quantum dots has been extensively investigated~\cite{OrellanaPRB73,PhysRevB.87.075102,TrochaPRB78,VernekPRB82}, and Fano--Andreev interference is also well established in hybrid nanostructures~\cite{LongBai_TQD_A,Siqueira3,CalleAM,CALLE20131474,Fano_Andreev_Effect_TQD,Dicke_Fano_Andreev,PhysRevB.95.125419}, their combined manifestation in an interacting finite-bias normal--superconducting triple-dot device has remained only partially understood. In particular, it remains unclear how lateral-dot detuning reshapes the effective coupling of the BIC-related subgap sector to the transport continuum in the presence of proximity-induced Andreev processes on the lateral dots. By employing the nonequilibrium Green's function method~\cite{keldysh2024diagram} (see Sec.~S1 of the Supplementary Material for details),  here we show that the detuning parameter provides direct control over this coupling and thereby drives a continuous crossover from a Fano--Andreev BIC-supported regime to a Fano--Andreev quasi-BIC regime. This crossover is reflected simultaneously in the evolution of finite ET antiresonances into exact transport zeros, the appearance of corresponding anomalies in the AR response, the broadening of narrow subgap spectral features, and changes in the lateral-dot occupation. Our results thus identify a concrete and experimentally accessible setting in which BIC formation, superconducting proximity, and interaction effects interplay and can be  probed on equal footing through hybrid mesoscopic transport.

\emph{Model.} We consider a triple-quantum-dot system in which the central dot \(QD_b\) is coupled to two normal leads, \(L\) and \(R\), while the lateral dots \(QD_a\) and \(QD_c\) are each coupled to superconducting electrodes, \(S_1\) and \(S_2\), respectively, as shown in Fig.~\ref{Fig1}. The Hamiltonian is
\begin{equation}
H = H_{L} + H_{R} + H_{S1} + H_{S2} + H_{TQD} + H_{T}.
\end{equation}

The normal (left and right) leads are described by
\begin{equation}
H_{\alpha}  =  \sum_{k,\sigma} \epsilon_{\alpha k \sigma}\, c^{\dagger}_{\alpha k \sigma} c_{\alpha k \sigma},
\end{equation}
where \(c^{\dagger}_{\alpha k \sigma}\) (\(c_{\alpha k \sigma}\)) creates (annihilates) an electron with spin \(\sigma\) and energy \(\epsilon_{\alpha k \sigma}\) in the \(\alpha=L,R\) electrode.

Each superconducting lead is modeled by a BCS Hamiltonian~\cite{BCS},
\begin{align}
&H_{S_{\alpha}}=  \sum_{k,\sigma} \epsilon_{kS_{\alpha}}\, c^{\dagger}_{k S_{\alpha} \sigma} c_{k S_{\alpha} \sigma} \nonumber\\
&+ \sum_{k} \left( \Delta_{\alpha}^{\ast} c_{k S_{\alpha} \downarrow} c_{-k S_{\alpha} \uparrow}
+ \Delta_{\alpha} c^{\dagger}_{-k S_{\alpha} \uparrow} c^{\dagger}_{k S_{\alpha} \downarrow} \right),
\end{align}
where \(c^{\dagger}_{k S_{\alpha} \sigma}\) (\(c_{k S_{\alpha} \sigma}\)) is the creation (annihilation) operator in superconducting electrode \(\alpha=1,2\), and \(\Delta_{\alpha}\) denotes the superconducting  gap of lead \(S_{\alpha}\), which we treat as real, that is, with a superconducting phase factor ($\phi_{\alpha}$) equal to zero.

The triple quantum dot is described by
\begin{align}
&H_{TQD} = \sum_{\sigma}\sum_{m=a,b,c} \epsilon_{m,\sigma}\, d^{\dagger}_{m \sigma} d_{m \sigma}\\
&+ \sum_{\sigma}\sum_{m=a,c} t_{mb}\left(d^{\dagger}_{m \sigma} d_{b \sigma} + \mathrm{H.c.}\right) + \sum_{m=a,b,c} U_m\, n_{m\uparrow} n_{m\downarrow},\nonumber
\end{align}
where \(d^{\dagger}_{m \sigma}\) (\(d_{m \sigma}\)) creates (annihilates) an electron with spin \(\sigma\) on dot \(m=a,b,c\). We assume a single spin-degenerate level on each dot, \(\epsilon_{m\uparrow}=\epsilon_{m\downarrow}\), where \(U_m\) is the intradot Coulomb interaction and \(t_{mb}\) the interdot hopping amplitude.

The coupling between the leads and the TQD is given by
\begin{align}
&H_T = \sum_{\beta=L,R}\sum_{k,\sigma}
\left( V_{k\beta}\, c^{\dagger}_{k \beta \sigma} d_{b \sigma} + \mathrm{H.c.} \right) \nonumber\\
&+ \sum_{k,\sigma} \left( V_{S_{1}k}\, c^{\dagger}_{S_{1} k \sigma} d_{a \sigma}
+ V_{S_{2}k}\, c^{\dagger}_{S_{2} k \sigma} d_{c \sigma} + \mathrm{H.c.} \right),
\end{align}
where \(V_{k\beta}\) is the tunneling amplitude between the central dot \(QD_b\) and the normal lead \(\beta=L,R\), while \(V_{kS_{\alpha}}\) describes tunneling between superconducting lead \(S_{\alpha}\) and the corresponding lateral dot. In the wide-band limit, the lead-induced broadening is
\begin{equation}
\Gamma_{\beta}=2\pi \sum_k |V_{k\beta}|^2 \delta(\omega-\epsilon_k^{\beta}).
\end{equation}
\emph{Results.} Hereafter, we discuss results obtained from the nonequilibrium Green's-function formalism, with local interactions treated within the Hubbard approximation~\cite{Meir1991,Qing-feng-Sun}. We use \(\Gamma_L\) as the unit of energy and  assume spin-degenerate dot levels with identical intradot Coulomb repulsion in all three dots, \(U_a=U_b=U_c=U\). The parameter \(\eta\) controls the detuning of the lateral-dot levels relative to \(\epsilon_b=\epsilon_d\), such that \(\epsilon_a=\epsilon_b+\eta\) and \(\epsilon_c=\epsilon_b-\eta\). Unless stated otherwise, the dot levels are set at the electron-hole symmetric point, \(\epsilon_a=\epsilon_b=\epsilon_c=-U/2\). For simplicity, we denote the electron-tunneling (ET) and Andreev-reflection (AR) differential conductances by \((dI/dV)^{\rm ET}\) and \((dI/dV)^{\rm AR}\), respectively. The normal leads are biased symmetrically, \(\mu_1=-\mu_2=eV\), while the superconducting leads remain grounded, \(\mu_S=0\).
In the parameter regime considered here, the relevant transport resonances lie inside the superconducting gap. Consequently, quasiparticle contributions from the continuum above the gap are strongly suppressed, and the low-energy transport is dominated by subgap Andreev processes and proximity-induced correlations. Our analysis is further restricted to the regime outside the Kondo limit. See Sec.~S2 of the Supplementary Material for details.

\begin{figure*}[ht]
    \centering
    \includegraphics[scale=0.5]{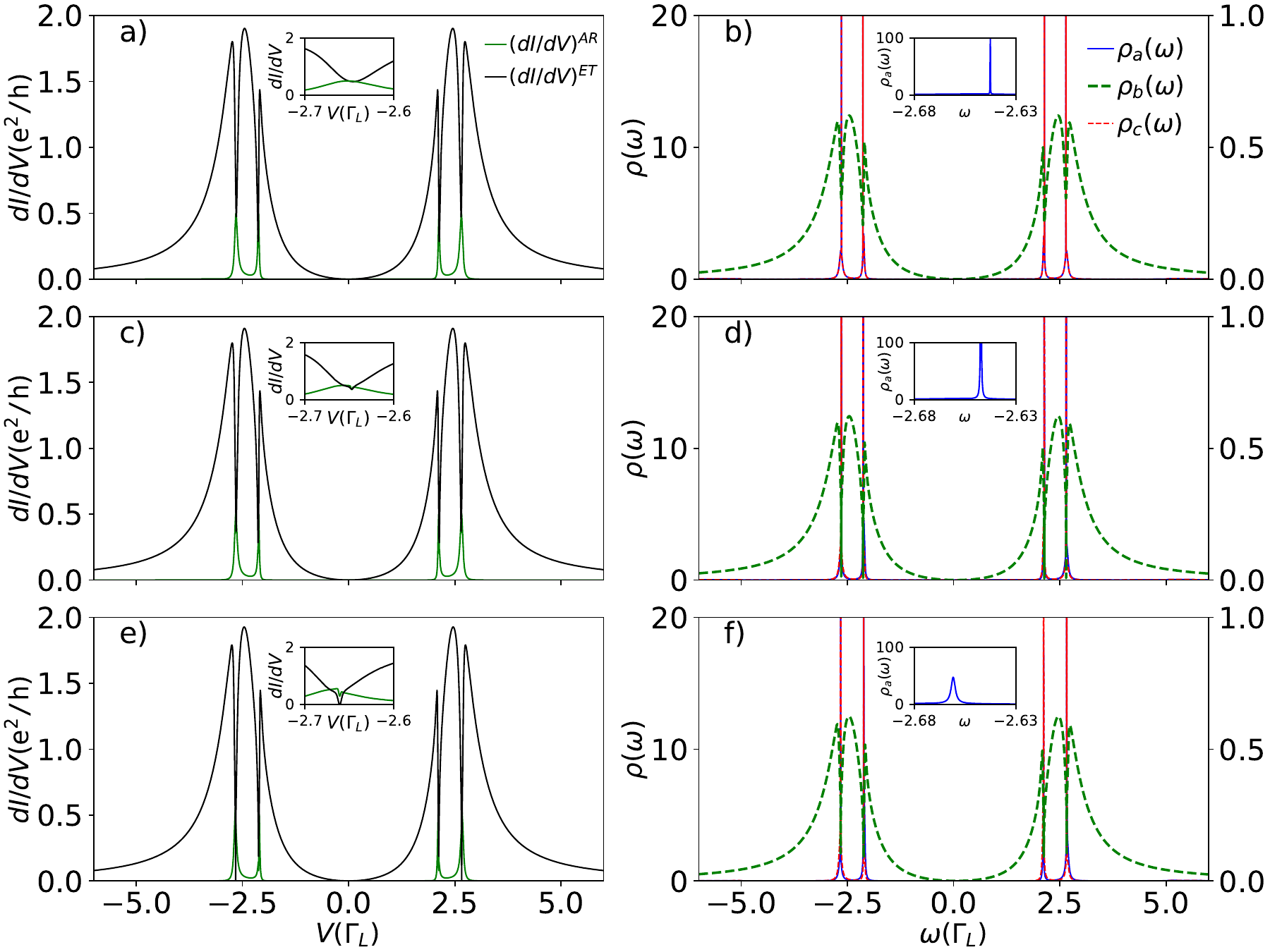}
    \caption{\justifying{Transport and spectral signatures of the interacting hybrid triple-quantum-dot system (Fig.~\ref{Fig1}) for fixed values of the lateral-dot detuning. Left column: Differential conductance for electron tunneling (ET), \((dI/dV)^{\rm ET}\), and Andreev reflection (AR), \((dI/dV)^{\rm AR}\), versus bias \(V\) for fixed value of lateral-dot detuning (a) \(\eta=0\), (c) \(\eta=0.05\), and (e) \(\eta=0.1\). Right column: corresponding local density of states (LDOS) \(\rho_i(\omega)\) as a function of energy ($\omega$) at quantum dots $i=a,b,c$, for  detuning (b) \(\eta=0\), (d) \(\eta=0.05\), and (f) \(\eta=0.1\), taken at the values of bias corresponding to the electron-tunneling antiresonances near \(V=-U/2\). Insets highlight the antiresonance region in the conductance (left column) and the narrow subgap peaks in the LDOS (right column). Other parameters are fixed to the following values:   \(\Delta_1=\Delta_2=5\Gamma_L\), \(U=5\), \(T=0\), \(\Gamma_L=\Gamma_R=\Gamma_S\), and \(\epsilon_{a,c}=\epsilon_b\pm\eta\) with \(\epsilon_b=-U/2\). See also Fig.~S1 in the Supplementary Material for the additional plots of the occupation number versus bias for the same values of the parameters.}}
    \label{figure2}
\end{figure*}

In Fig.~\ref{figure2}(a)--(c), we show the interacting ET [\((dI/dV)^{\rm ET}\), solid lines] and AR [\((dI/dV)^{\rm AR}\), dashed lines] differential conductances as functions of the bias \(V\) for several values of the detuning \(\eta\), defined by \(\varepsilon_{a,c}=\varepsilon_b\pm\eta\). The derivation of \((dI/dV)^{\rm ET}\) and \((dI/dV)^{\rm AR}\) is given in Sec.~II of the Supplementary Material. For \(\eta=0\) and \(t_{ab}=t_{cb}\) [Fig.~\ref{figure2}(a)], \((dI/dV)^{\rm ET}\) exhibits finite-depth antiresonances on both sides of the Hubbard poles \((\epsilon=\pm U/2)\), which lie inside the superconducting gap, while \((dI/dV)^{\rm AR}\) shows corresponding peaks. In this symmetric limit, the two lateral dots form symmetric and antisymmetric combinations, \( |+\rangle=(|a\rangle+|c\rangle)/\sqrt{2}\) and \( |-\rangle=(|a\rangle-|c\rangle)/\sqrt{2}\). The symmetric sector couples directly to the central dot and hybridizes with both the normal continuum and the superconducting leads. By contrast, the antisymmetric sector decouples from the central dot by destructive interference between the two lateral paths and therefore does not participate directly in normal transport. Within the superconducting gap, the BCS density of states vanishes and the superconducting self-energy is purely real, so the superconducting electrodes do not provide an additional dissipative channel. As a result, the effective decay rate of the antisymmetric sector vanishes, yielding a BIC with real energy embedded in the metallic continuum but infinite lifetime with respect to decay into the normal leads. {In this sense, the decoupled antisymmetric state is not merely an electronic dark state. Since it lies in the superconducting subgap regime and is dressed by the proximity-induced anomalous self-energies, it can be interpreted more precisely as an Andreev bound state in the continuum}.

At the same time, superconducting proximity induces pairing on the lateral dots and generates ABS inside the gap, which opens virtual paths between the lateral subsystem and the central transport channel. The interference between the direct elastic path through the central dot and the indirect ABS-mediated path then produces the asymmetric Fano-like structure in \((dI/dV)^{\rm ET}\), with a complex asymmetry parameter \(q\), and hence finite-depth dips rather than exact zeros. These features are therefore not due to a purely Fano mechanism; instead, they reflect the combined action of the BIC-induced reorganization of the lateral subsystem into symmetric and antisymmetric sectors and Fano--Andreev interference in the transport channel. The former yields a narrow long-lived subgap state with weak effective coupling to the continuum, while the latter converts it into an observable antiresonance through interference between resonant and background transmission paths.

This interpretation is consistent with Fig.~\ref{figure2}(b), where for \(\eta=0\) the local density of states (LDOS) of the lateral dots, \(\rho_{a/c}\), develops extremely sharp subgap peaks with only weak broadening. These peaks indicate symmetry-protected long-lived states and are consistent with a BIC-like configuration. Still, an almost \(\delta\)-like LDOS peak does not by itself establish a true BIC; rather, it signals a very small effective decay rate. A genuine BIC requires the effective coupling to the continuum to vanish, \(\Gamma_{\rm eff}=0\), which in the symmetric limit is more naturally associated with the antisymmetric sector than with the LDOS of either lateral dot separately. Indeed, \(\rho_a(\omega)\) and \(\rho_c(\omega)\) contain contributions from both the symmetric \((|+\rangle)\) and antisymmetric \((|-\rangle)\) sectors. The observed peaks should therefore be understood as superconductivity-dressed subgap resonances with very long lifetime, whose residual width is set by their weak hybridization with the transport channel.

A useful check is obtained by setting \(\Delta=0\), so that \(S_1\) and \(S_2\) become normal electrodes. In that limit, the dips in \((dI/dV)^{\rm ET}\) disappear. This shows that the observed antiresonances are of genuine Fano--Andreev origin: without superconducting pairing, the indirect amplitude mediated by the lateral subsystem lacks the Andreev-dressed resonant structure required to cancel the direct central-dot contribution. Once superconducting proximity is restored, the indirect transport pathway is qualitatively reshaped and interferes destructively with the background ET contribution, thereby producing the Fano-like conductance dips. Since the Hubbard poles remain inside the superconducting gap, the complex character of the Fano parameter \(q\) is not inherited directly from the superconducting leads, but indirectly from the normal-lead self-energy transmitted through the lateral-dot sector, which supplies the dissipative phase of the transport channel.

Figures~\ref{figure2}(c) and \ref{figure2}(e) show that, once a small but finite detuning \(\eta\neq0\) is introduced, the antiresonances in \((dI/dV)^{\rm ET}\) deepen and eventually become exact zeros at finite bias \(V\simeq \pm U/2\), while Figs.~\ref{figure2}(d) and \ref{figure2}(f) display the corresponding spectral evolution in the LDOS. Simultaneously, \((dI/dV)^{\rm AR}\) develops pronounced minima, indicating that Andreev transport is suppressed by the same destructive-interference mechanism. Physically, finite detuning mixes the symmetric and antisymmetric sectors, so that the antisymmetric sector acquires a finite coupling to the transport continuum through the central dot. In the presence of superconducting proximity, the ABS on the lateral dots provide additional virtual tunneling paths, allowing this weakly coupled subgap sector to participate actively in transport interference. The coexistence of these channels permits complete destructive interference between direct and indirect processes, yielding exact zeros in \((dI/dV)^{\rm ET}\) at specific bias values.

This evolution is also reflected in the LDOS [see Fig.~\ref{figure2}(d) and (f)], where the lateral-dot peaks broaden progressively with increasing \(\eta\). This \(\eta\)-dependent broadening shows that detuning weakens the interference condition protecting the long-lived states in the nearly symmetric regime. Accordingly, the antisymmetric sector hybridizes more strongly with the symmetric transport-active sector and, through the central dot, with the normal continuum. Its effective decay rate ( \(\Gamma_{\rm eff}\) ) therefore increases with \(\eta\), and the LDOS evolves from an almost \(\delta\)-like structure to a finite-width resonance. Because the lateral dots are not directly coupled to the normal leads, this broadening should be understood as an indirect leakage process mediated by the central dot. The \(\eta\)-dependence thus provides direct spectral evidence for a crossover from a BIC regime to a quasi-BIC regime. In parallel, the AR conductance develops antiresonances at bias values close to, although not exactly coincident with, those of \((dI/dV)^{\rm ET}\). This shows that the same quasi-BIC that suppresses normal transmission also weakens the coherent electron-hole conversion responsible for Andreev reflection. Both features originate from the same detuning-controlled interference mechanism, although they involve different components of the Green's function: the normal component for ET and the anomalous one for AR. We therefore identify a continuous crossover from a Fano--Andreev BIC-supported regime to a Fano--Andreev quasi-BIC regime as \(\eta\) increases.

Figures~3(a) and 3(b) show contour plots of \((dI/dV)^{\rm ET}\) and \((dI/dV)^{\rm AR}\) versus \(\eta\) and \(V\) for the particle--hole symmetric case, \(\epsilon_b=-U/2\). Consistent with the discussion above, for \(\eta=0\), \((dI/dV)^{\rm ET}\) displays finite-depth antiresonances that do not reach zero, while \((dI/dV)^{\rm AR}\) exhibits corresponding peaks. As \(\eta\) increases, for instance to \(\eta=0.1\), the ET antiresonances deepen and eventually become exact zeros at finite bias, \(V\simeq \pm U/2\), whereas the AR conductance develops pronounced minima.The same trend is further illustrated in Fig.~S3 of the Supplementary Material, where \((dI/dV)^{\rm ET}\) and \((dI/dV)^{\rm AR}\) are shown over a wider bias range for representative values of \(\eta\). Figure~S4 provides the complementary linear-response behavior, displaying \(G^{\rm ET}\) and \(G^{\rm AR}\) as functions of the central-dot level \(\epsilon_b\). Overall, the \(\eta\)-dependence of the differential conductance provides a clear signature of the BIC-to-quasi-BIC crossover: increasing \(\eta\) enhances the mixing between the symmetric and antisymmetric sectors. In the ET sector, this drives a crossover from a Fano--Andreev BIC-supported regime at \(\eta=0\) to a Fano--Andreev quasi-BIC regime at finite \(\eta\), where the minima arise from destructive interference between the background transport channel and a narrow subgap resonance. By contrast, the AR response follows more directly the Andreev spectral structure, showing resonant bands with weaker antiresonant suppression. Taken together, these maps indicate that \(\eta\) primarily controls the effective decay rate of the quasi-BIC sector, rather than the position of the subgap resonances.

\begin{figure}[t!]
\includegraphics[scale= 0.45]{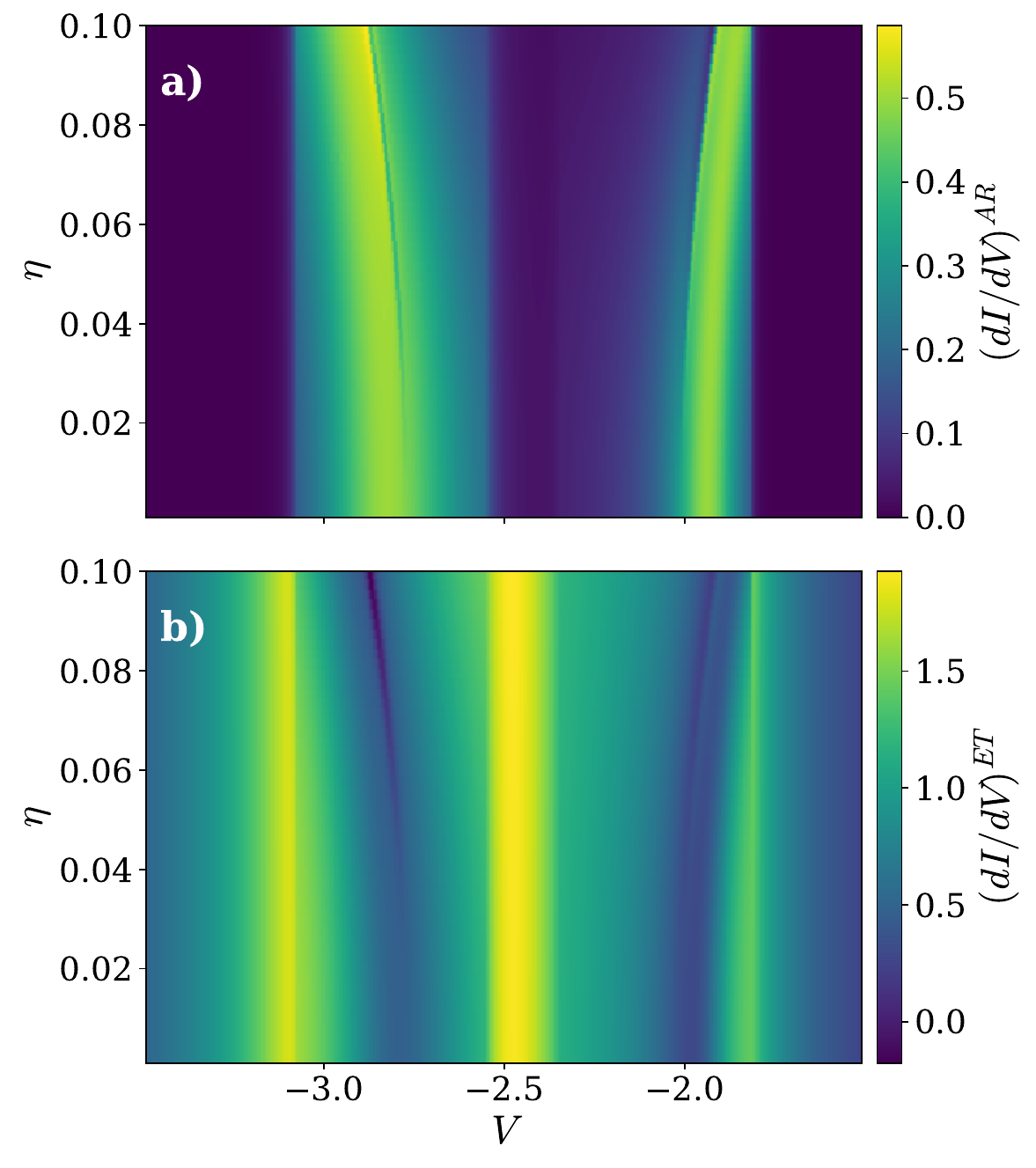}
\caption{\justifying{
Contour plots of the differential conductances \((dI/dV)^{\rm ET}\) and \((dI/dV)^{\rm AR}\) versus the lateral-dot detuning \(\eta\) and bias \(V\), for the interacting hybrid triple-quantum-dot system (Fig.~\ref{Fig1}). Here, ET and AR denote electron tunneling and Andreev reflection, respectively: (a) \((dI/dV)^{\rm ET}\) and (b) \((dI/dV)^{\rm AR}\). As \(\eta\) increases, the ET antiresonances deepen and reach exact zeros near \(V\simeq \pm U/2\), while the AR conductance develops corresponding minima. Other parameters are chosen as \(\Delta_1=\Delta_2=5\Gamma_L\), \(U=5\), \(k_B T=0\), \(\Gamma_L=\Gamma_R=\Gamma_S\), and \(\epsilon_{a,c}=\epsilon_b\pm\eta\) with \(\epsilon_b=-U/2\). See Fig.~S2 in the Supplementary Material for the corresponding plots in a wider range of $V$. }}
\label{figure4}
\end{figure}

\emph{Summary and discussion.} In this work, we analyzed quantum transport in an interacting triple quantum dot system coupled to normal and superconducting electrodes, with particular emphasis on the emergence and evolution of BICs driven by interference effects.

Our results highlight the interplay of Coulomb interactions, superconducting proximity, and quantum interference in hybrid nanostructures. They provide a unified interpretation of the BIC mechanism in the interacting triple-dot subsystem and its interplay with the proximity-induced Fano--Andreev effect, and establish these devices as a versatile platform for engineering and controlling BICs in mesoscopic superconducting systems. In particular, we identify a continuous crossover from a Fano--Andreev BIC-supported regime to a Fano--Andreev quasi-BIC regime as the detuning parameter \(\eta\) increases. In the symmetric limit \((\eta=0)\), the antisymmetric sector is fully decoupled from the transport channel. {Because this decoupled sector remains in the subgap window and is induced by superconducting proximity, it can be viewed, in the ideal symmetric limit, as an Andreev bound state in the continuum.}. With increasing detuning, destructive interference in the elastic-tunneling channel is reinforced, driving the antiresonances to exact zeros and producing quasi-BIC behavior. Simultaneously, the Andreev differential conductance develops corresponding minima, reflecting a redistribution of spectral weight between elastic and Andreev processes. We further show that, in the interaction-driven regime, the emergence of quasi-BIC and BIC features is intrinsically tied to bias-induced charge redistribution on the lateral dots, rather than being a purely single-particle interference effect.


In conclusion, our results reveal clear and experimentally accessible signatures of the detuning-driven crossover from a Fano--Andreev BIC-supported regime to a Fano--Andreev quasi-BIC regime. These signatures can be directly probed by tunneling spectroscopy, in particular through differential-conductance \((dI/dV)\) measurements and local spectroscopic probes capable of resolving the narrow subgap resonances. Experimentally, the key control parameter is the lateral-dot detuning \(\eta\), which can be tuned by independent gate voltages applied to the side quantum dots~\cite{Granger2010,Burkard2023}. This gate control provides a direct route to track the progressive evolution from finite ET antiresonances to exact transport zeros, together with the accompanying changes in the AR response and in the spectral width of the lateral-dot resonances. 

More broadly, our findings show that hybrid triple-quantum-dot devices offer a realistic platform in which interference effects, superconducting proximity, and local Coulomb interactions can be manipulated on equal footing. In this sense, the present setup provides a concrete framework for investigating how long-lived BIC-related subgap states emerge, hybridize, and eventually leak into the transport continuum. We therefore expect our results to be relevant not only for the identification of BIC and quasi-BIC features in mesoscopic superconducting transport, but also for the broader design of hybrid nanostructures in which interference can be engineered and electrically tuned with high precision.

\emph{Acknowledgments.} P.A.O. acknowledges support from DGIIE USM PI-LIR-24-10, FONDECYT Grants No. 122070 and 1230933. V. J.  acknowledges support from  FONDECYT Grant No.~1230933. A. G. acknowledges support from  FONDECYT Grant No.~3250671.

\bibliography{draft_references_paper_BICs}

\end{document}